\documentclass[10pt]{article} 
\usepackage{amssymb}
\usepackage{latexsym} 
\usepackage[dvips]{epsfig}

\def\d{\mbox{d}}

\font\tenscr=rsfs10 scaled1100
\font\sevenscr=rsfs7 
\font\fivescr=rsfs5 
\skewchar\tenscr='177 
\skewchar\sevenscr='177
\skewchar\fivescr='177 
\newfam\scrfam 
\textfont\scrfam=\tenscr
\scriptfont\scrfam=\sevenscr 
\scriptscriptfont\scrfam=\fivescr

\def\scri{{\fam\scrfam I}}

\begin{document}

\title{Can one detect a non-smooth null infinity?}  
\author{Juan Antonio Valiente Kroon \thanks{E-mail address:
{\tt jav@aei-potsdam.mpg.de}} \\
Max Planck Institut f\"ur Gravitationsphysik,\\ Albert Einstein Institut,\\
Am Muhlemberg 1, 14476 Golm,\\ Germany.}

\maketitle

\begin{abstract}
It is shown that the precession of a gyroscope can be used to elucidate the
nature of the smoothness of the null infinity of an asymptotically flat 
spacetime (describing an isolated body). A model for which the effects 
of precession in the non-smooth null infinity case are of order $r^{-2}\ln r$ 
is proposed. By contrast, in the smooth version the effects are of order 
$r^{-3}$. This difference should provide an effective criterion to decide 
on the nature of the smoothness of null infinity.  

\end{abstract}

\section{Introduction}
For sometime now, the smoothness or not of null infinity of an 
asymptotically flat spacetime has remained as an issue of debate in the 
mathematical studies of the Einstein field equations.
\emph{La raison d'\^{e}tre} of the asymptotically flat spacetimes is to 
provide a model to describe the gravitational field of isolated bodies within 
the framework of General Relativity. The way one could define an isolated body
 in GR has been the source of much, and in a way still present debate (see 
for example \cite{Ell84}). The most
 favoured solution was put forward by Penrose \cite{Pen63,Pen65a}, and consists
 of performing a conformal rescaling of the spacetime which results in the 
attachment of a boundary to the original manifold. Losely speaking, one will 
have a spacetime describing an isolated body, whenever such a 
``compactification procedure'' can be performed, and the associated boundary 
(null infinity, $\scri$ \footnote{The symbol $\scri$ denoting
null infinity is called ``scri''. Incidentally, the phonetic transliteration 
of scri into Polish is \emph{skraj}, which curiously enough means boundary!}) 
is a null hypersurface. 

An outsider could regard the considerations on the smoothness of null 
infinity as just mere technical matters without real physical significance, 
beyond allowing the mathematical treatment of the problem. It is the 
objective of this article to point out that there are physical effects 
differentiating a spacetime with a smooth null infinity and one with a 
non-smooth one. 

In order to make the discussion more precise, the forthcoming analysis 
will be centered in the class of asymptotically flat spacetimes with a 
non-smooth null infinity known as polyhomogeneous spacetimes. These spacetimes 
possess the peculiarity of having asymptotic expansions in terms of a 
parameter $r$ and its logarithm, $\ln r$. The presence of logarithms in 
the asymptotic expansions yields as a result a non-smooth null infinity 
\cite{ChrMacSin95}. More recently, work on a new representation of 
spatial infinity has shown the appearance of logarithmic singularities at
``the points where null infinity meets spatial infinity'' \cite{Fri98a}
 These logarithmic terms have a long history, and can 
also appear in similar expansion for linear fields\cite{Val00a}. As early 
as the late 50's independent work by Fock and Bonnor contained such 
logarithms. Some years later the seminal work by Sachs \cite{Sac61,Sac62b} 
and Bondi et al. \cite{BonBurMet62} invoked an ``outgoing radiation 
condition'' analogous to that of Sommerfeld (see \cite{Som92} and also 
\cite{CouHil62}) for the wave equation and  
the electromagnetic field in order to preclude the appearance of logarithms 
in the asymptotic expansions. There has been some confusion in the literature
 about the meaning and significance of these kind of radiation conditions. 
Sommerfeld's condition for the wave equation in Minkowski spacetime aims 
to exclude incoming radiation, and not to ensure the existence of outgoing 
radiation. Since the time this condition was formulated, it has been noticed 
that retarded solutions for spatially bounded sources satisfy Sommerfeld's 
condition automatically in future directions. However, these solutions do 
not necessarily satisfy these conditions in past null directions. If one 
wants to exclude incoming radiation, conditions have to be imposed
precisely along the past null directions coming from past null infinity 
($\scri^-$). As one recedes towards past null infinity along null curves, 
one finds  that the retarded field reflects the source behaviour at earlier 
times. Thus, a condition on the time dependence of the source in the infinite 
past is required so that the retarded field satisfies Sommerfeld's 
condition at $\scri^-$ \cite{Wal79}. Conditions of this type have long time 
been given for the wave equation, the gravitational field and linearised 
gravity \cite{LeiWal77}. In the case of the full gravitational field, the 
situation is different because there are no integral representations of the 
field. The presence of incoming radiation at any time cannot be avoided 
(in a radiative spacetime, of course!), and can be interpreted as describing 
the phenomena of gravitational wave scattering and gravitational wave tails 
that die off suitably in a neighbourhood of null infinity.

Spacetimes with a smooth null infinity on the one hand,  and
polyhomogeneous spacetimes on the other, 
are two classes of solutions of the Einstein field equation which
attempt to provide a description of the physics of isolated
bodies. Now, given a particular physical system (which can be
approximately considered as an isolated system), to which class should belong
the solution describing it? A further related question is whether
these classes are large enough to describe most systems of physical
interest, or do we need to look for more general families of
solutions? The answer to these questions must come through the
comparison of the predictions given by each type of solution with the
observations. In particular, one would like to have some specific
physical effect which could be use as a clear ``finger print''of a
particular class of solutions.


One possible way of detecting the presence of gravitational radiation is 
through its effects on a gyroscope. Herrera \& Hern\'{a}ndez \cite{HerHer00} 
have discussed how to extract information about gravitational waves 
by means of their effects on a gyroscope moving on a particular world line. 
Here, these ideas are taken over in order to show that in principle it is
possible to make some statements on the nature of the null infinity of the
spacetime. In order to do so, a simple (oscillatory) model for the 
news function of a radiative system is put forward. Under this assumption
it is shown that the effects of precession of a gyroscope in a polyhomogeneous 
spacetime can be an order of magnitude stronger than those appearing in the
peeling counterpart. This effect could be used to decide whether a radiative 
system should be modelled by means of a smooth or a non-smooth null infinity.

\section{The Bondi metric}
The early studies of the asymptotic behaviour of the gravitational 
radiation produced by an isolated body were carried out by means 
of the construction of an \emph{ad hoc} metric and associated coordinate 
system \cite{BonBurMet62}. The referred metric is now widely known as the 
axially symmetric Bondi metric. In the most standard parametrisation it 
reads as follows:

\begin{equation}
\d s^2= \left( \frac{V}{r}e^{2\beta}-U^2r^2e^{2\gamma}\right)\d u^2 
+2e^{2\beta}\d u \d r + 2Ur^2e^{2\gamma}\d u\d \theta - r^2\left(e^{2\gamma}
\d\theta^2+e^{-2\gamma}\sin^2\theta\d \varphi^2 \right). \label{Bondimetric}
\end{equation} 
The coordinate $u$ is a retarded time labelling (future oriented) light cones, 
and $(\theta,\varphi)$ are the usual angular coordinates of the celestial 
sphere. The crucial ingredient of the Bondi metric is the coordinate $r$, 
usually known as \cal{luminosity parameter}. It satisfies
\begin{equation}
r^4\sin^2\theta=\det h_{ij},
\end{equation}
where $h_{ij}$ is the angular part of the metric. This metric assumes the 
existence of an hypersurface orthogonal axial Killing vector field. In 
addition, the metric functions are required to satisfy some regularity 
conditions on the poles. But these, shall not concern us here. The 
generalisation of eqn.(\ref{Bondimetric}) to the case where no symmetries 
are present was studied by Sachs \cite{Sac62}. 

The present analysis will be restricted to the axial symmetric case, 
mainly in order to ease the calculations. Nevertheless, as pointed out by 
Sachs himself \cite{Sac62}, no interesting physics lost with this 
assumption as all the relevant features of the gravitational radiation 
phenomena are generally already present in the axially symmetric case.

\section{Polyhomogeneous spacetimes}

The asymptotic expansions of the metric coefficients of the Bondi metric for 
the case of a peeling asymptotically flat spacetime are very well known. 
Their leading terms read:
\begin{eqnarray}
&&\gamma_{*}=c_0r^{-1}+O(r^{-3}), \\
&&\beta_{*}=-\frac{1}{4}c_0^2r^{-2}+O(r^{-4}), \\
&&U_{*}=-(c_{0\theta}+2c_0\cot\theta)r^{-2}+O(r^{-3}), \\
&&V_{*}=r-2M+O(r^{-1}).
\end{eqnarray}
Note here the absence of the $r^{_2}$-term in the expansion of $\gamma$. 
This is the way the ``outgoing radiation condition'' (cfr. the introduction) 
is imposed. 

The most natural and simple example of a polyhomogeneous spacetime is the 
so-called minimally polyhomogeneous for which \footnote{The subindex ${}_\#$
will be used to denote functions with polyhomogeneous expansions, while 
${}_*$ will be used to denote functions with analytic expansions in $1/r$.}:
\begin{equation}
\gamma_{\#}=cr^{-1}+ \gamma_2r^{-2}+O(r^{-3}\ln r).
\end{equation}
The presence of the extra term $\gamma_2$ in the expansion gives rise to 
logarithmic terms at higher order $O(r^{-3})$ in $\gamma$ and at order 
$O(r^{-4})$, $O(r^{-3})$, and $O(r^{-1})$ in the metric functions $\beta$, 
$U$, and $V$ respectively. Looking at these spacetimes in the Newman-Penrose
framework one sees that $\Psi_0=O(r^{-4})$, i.e. these 
spacetimes are non-peeling \cite{Val99a}. In the present article, a more general type of 
polyhomogeneous spacetime will be used. Namely that one generated by a metric 
function $\gamma$ of the form:
\begin{equation}
\gamma_{\#}= \left(c_1\ln r +c_2\right)r^{-1}+\left(\gamma_{21}\ln r + 
\gamma_{20}\right)r^{-2}+O(r^{-3}\ln r).
\end{equation}
In this case, the leading terms of the remaining metric functions are found to be:
\begin{eqnarray}
&&\!\!\!\!\!\!\!\!\beta_{\#}=\frac{1}{4}\left( -c_1^2\ln^2 r+ 
\left(c_1^2
-2c_1c_0\right)\ln r -c_0^2+c_0^2+
c_1c_0-\frac{1}{2}c_1^2\right)r^{-2}+O(r^{-3}\ln^2r), \\
&&\!\!\!\!\!\!\!\!U_{\#}=\left( \left(-c_{1\theta}-2\cot\theta c_1\right)
\ln r 
-c_{0\theta}+\frac{3}{2}c_{1\theta}-2\cot\theta c_0+3\cot \theta c_1 
\right)r^{-2}+ O(r^{-3}\ln r), \\
&&\!\!\!\!\!\!\!\!V_{\#}= r-2M +\left(c_{1\theta\theta}+3\cot\theta 
c_{1\theta}
-2c_1\right)+O(r^{-1}\ln r). 
\end{eqnarray}
It can be shown that the coefficient $c_1$ is a constant of motion, 
i.e. $\partial_u c_1=0$ \cite{ChrMacSin95,Val99b}. This  more general 
class of spacetimes is such that 
$\Psi_0=O(r^{-3})$, and remarkably
does possess a well defined Bondi mass \cite{ChrJezMac98a,ChrJezMac98b}.

\section{A model of a gyroscope}
Consider an observer with world line given by $r=const$, $\theta=const$, 
$\varphi=const$. This observer has a 4-velocity given by:
\begin{equation}
u_a=\left(A,e^{2\beta}A^{-1},Ur^2e^{2\gamma}A^{-1},0\right),
\end{equation}
with
\begin{equation}
A=\sqrt{Ve^{2\beta}r^{-1}-U^2r^2e^{2\gamma}}.
\end{equation}
The precession of a gyroscope moving along such a world line is quantified 
through the vorticity of a congruence of world lines surrounding our 
fiduciary world line (for a discussion on the description of precession 
effects we refer to \cite{RinPer90}, and more recently to \cite{HerMarRui00}). The vorticity vector is given by:
\begin{equation}
\omega^a=\frac{1}{2\sqrt{-g}}\varepsilon^{abcd}u_bu_{c,d}.
\end{equation}
Thus, for the Bondi metric one has that \cite{HerHer00}:
\begin{eqnarray}
&&\Omega=\sqrt{-\omega_a\omega^a} \nonumber \\ 
&&\;\;\;\; =\frac{1}{2}r^{-1}e^{-2\beta-\gamma}\left[ 2\beta_\theta e^{2\beta}-2e^{2\beta}A_\theta A^{-1}-(Ur^2e^{2\gamma})_r+2Ur^2e^{2\gamma}A_rA^{-1} \nonumber \right. \\
&&\;\;\;\;\;\;\;\left.+e^{2\beta}(Ur^2e^{2\gamma})_u-2\beta_ue^{2\beta+2\gamma}Ur^2A^{-2}\right]. \label{vorticity}
\end{eqnarray}
The scalar $\Omega$ measures the rate of rotation with respect to the proper 
time of world lines of points with $r=const.$, $\theta=const.$, 
$\varphi=const.$. Thus, it describes the precession of a gyroscope moving 
along the fiduciary world line.

For a peeling spacetime one has that $\Omega$ is given by,
\begin{equation}
\Omega_{*}=-\frac{1}{2r}\left(\partial_u\partial_\theta c + \cot\theta\partial_u c\right)+ O(r^{-2}).
\end{equation}

\section{A model to study precession}

The effect of a gravitational wave on a ring of test particles in a plane perpendicular to the direction of propagation of the wave is well known. It can be deduced from the geodesic deviation equation. In this case it reads:

\begin{equation}
\delta\ddot{x}^a=-\frac{1}{2}\ddot{c_0}\left(e^a_2e_{2c}-e_1^ae_{1c}\right)\delta x^c
\end{equation}
where $e^a_1$ is an azimuthal vector, and $e^a_2$ is a polar one. Due to the axial symmetry, the North Pole and the South Pole of the Celestial sphere are already fixed. Furthermore, due to the orthogonal transitivity of the axial Killing vector, there is only $(+)$-polarisation (the axes of stretching and contraction lie along the $e^a_1$ and $e^a_2$ directions).

In order to study the possible effects of the gravitational radiation on a gyroscope, the following model for the expansion coefficient $c_0$ is put forward:

\begin{equation}
c_0=\frac{k}{\omega^2}\sin (\omega u),
\end{equation}
where $k$ is a constant, and $\omega$ is an arbitrary period. Thus, this model
 produces periodic oscillatory distortions of our ring of test particles. 
Its derivative yields the news function. This particular form of the news 
function will be used for both the peeling and polyhomogeneous cases. It 
is very important to point that this kind of periodic behaviour is an 
approximation that can only be valid for a limited period of retarded time. 
The radiative system should settle to a quiescent stationary state as one 
approaches future time infinity. In other words, there are no exactly 
periodic radiative spacetimes.

For concreteness, we will only be interested in analyzing what happens at 
the equatorial plane, i.e. $\theta=\pi/2$. Thus, it will not worry us that 
the function $c_0$ here suggested does not satisfy the regularity 
requirements at the poles.

Substitution into \ref{vorticity} and evaluating at the equatorial plane 
one gets:

\begin{equation}
\Omega_{*}= \partial_\theta M r^{-2} + O(r^{-3}),
\end{equation}
that is, there are no $1/r$ precession effects. Those at order $1/r^2$ 
depend only the ``polar asymmetry'' of the mass aspect of the spacetime. 
The mass
aspect, M, is related to the news function via:
\begin{equation}
M_u=-c_{0u}+\frac{1}{2}\left(c_{0\theta\theta}+3c_{0\theta}\cot \theta 
-2c_{0}\right)_u,
\end{equation}
This equation holds for both the peeling and the polyhomogeneous spacetimes. 
Thus one has that
\begin{equation}
M=-\frac{uk^2}{2\omega^2}-\frac{k}{\omega}\sin(\omega u)-\frac{k^2}
{\omega^3}\sin(2\omega u)+f(\theta),
\end{equation}
where $f(\theta)$ is an arbitrary function coming out of the integration. 
If one particularises the model further more by requiring 
$\partial_\theta M=0$, i.e. $f(\theta)$, then one arrives to a 
situation where there are no $1/r^2$ effects!
      
Now, in the case of a polyhomogeneous $\scri$ one gets,

\begin{equation}
\Omega_{\#}=\left( \frac{5}{3} \partial_{\theta} c_1 - \frac{1}{3} 
\partial_{\theta\theta\theta}c_1 +\frac{1}{6}\partial_u c_0 
\partial_{\theta} c_1 +\partial_{\theta\theta} \right) r^{-2}\ln r 
+ O(r^{-2}).
\end{equation}
Hence, there is an $r^{-2}\ln r$ order precession effect on a gyroscope 
moving along the fiduciary world line, while the effects in a peeling 
spacetime appear at order $r^{-2}$ (or $r^{-3}$ if the mass aspect happens
to be isotropic!). This provides an effective criterion to differentiate
the two classes of null infinity.

\section{Conclusions}
It has been shown that the precession of a gyroscope can be used to 
investigate 
the nature of the null infinity modelling an isolated gravitational system. 
The analysed model is quite simple, but nevertheless it seems to shed some 
light on this issue. And therefore providing some physical interpretation 
to some issues that may seem of a purely mathematical nature. 

There has been some discussion on the nature of the physical interpretation 
of the logarithmic terms appearing in the expansions of asymptotically
flat spacetimes. In \cite{Val98} these terms have been regarded as associated
to incoming radiation and wave tails. Herrera \& Hern\'{a}ndez \cite{HerHer00}
 have argued that the $1/r^2$ effects are somehow related to incoming 
radiation in the form of wave tails. These ideas seem to bring further 
support to the relation between incoming radiation and logarithmic terms
in asymptotic expansions.

\section*{Acknowledgements}
Thanks to Dr. R. Lazkoz for reading a previous version of this essay. The note
on the meaning of scri in the Polish language is due to Dr. J. Jezierski. 
I also thank Dr. L. Herrera for providing a version of their work prior to
publication. Finally, an anonimous referee is thanked for constructive
criticism.


\end{document}